\documentclass[utf8]{frontiersSCNS} 

\usepackage{url,hyperref,lineno,microtype,subcaption}
\usepackage[onehalfspacing]{setspace}
\usepackage{mathptmx}
\usepackage{helvet}
\usepackage{courier}
\usepackage{type1cm}         

\usepackage{makeidx}         
\usepackage{graphicx}        
\usepackage{multicol}        
\usepackage[bottom]{footmisc}

\usepackage{mathtools}
\usepackage{amsmath}

\def\Authors{Roshan Gopalakrishnan}
\def\Address{Institute for Infocomm Research (I2R), ASTAR, Singapore}

\begin{document}
\onecolumn
\firstpage{1}

\title{RRAM based neuromorphic algorithms} 

\author{\Authors}
\address{} 

\extraAuth{}

\maketitle

\section{Introduction}

Human brain performs massively parallel and low power operations. It can outperform present age microprocessors on many tasks involving pattern recognition and input classification. The underlying neurons are heavily inter-connected; on average each neuron is connected to 10,000 (or up to 100,000) other neurons \cite{book_koch, AIP, Laughlin}. Despite the complexity of the human brain, research to understand the human brain is on-going, with the hope of emulating it in terms of its functionalities.

In recent years, RRAM devices have emerged as a major memory component in mimicking the functionality of synapses in the human brain\cite{Indiveri_2016}. This is mainly because RRAM can be used both as a memory element and computation unit. As mentioned in \cite{Cao_2015}, there are two ways of looking at RRAM based neuromorphic algorithms. From the deep learning perspective, one is to design algorithms for inference only, i.e to map the pre-trained deep learning models which fulfil certain hardware constraints onto the RRAM based neuromorphic hardware without any further training. While another way is to perform on-chip training on the RRAM based neuromorphic hardware, which will require additional interface circuitry for specific algorithms. Inference alone requires the conversion of existing pre-trained deep learning algorithms in high precision digital domain to the binary event-based (or spiking) domain so as to be able to be mapped onto RRAM based neuromorphic hardware. Whereas, on-chip training may be implemented at the RRAM synapse in the neuromorphic hardware by emulating local spike timing based algorithms such as spike timing dependent plasticity or its variants. These two methods belong to a new computational paradigm known as spiking deep neural network (SDNN).

Other than the aforementioned learning algorithms that can be implemented on RRAM based neuromorphic hardware, low precision convolutional neural networks (CNN) such as the binarized neural network \cite{BNN}, binaryNet \cite{BinaryNet}, XNOR-NET \cite{XNORNet} and DoReFa-NET \cite{DoReFaNet} can be mapped onto a chip containing RRAM based synaptic crossbar array \cite{Yu_Hao_RRAM}. In such an approach, the computations performed in the CNN maybe converted to bitwise operations, such as bitwise convolution, batch normalization and pooling etc., as shown in figure 1 of \cite{Yu_Hao_RRAM}. 
Contrary to other paradigms, mapping is much easier with such an approach as it does not involve spiking neurons. Irrespective of the mapping algorithms implemented on the RRAM based neuromorphic hardware, we should expect a drop in accuracy due to hardware noise, especially the noise inherent in RRAM synapses (Set or reset variability \cite{Stefano_set_reset}, Random Telegraph Noise (RTN) \cite{Stefano_RTN} etc.). One plausible approach to mitigate the drop in accuracy is to account for the noise itself during training, which may help to alleviate the accuracy loss to some extent.

\subsection{RRAM Synapse}
\label{sec:1.3.1}

\begin{figure}[htbp]
\centerline{\includegraphics[scale=0.7]{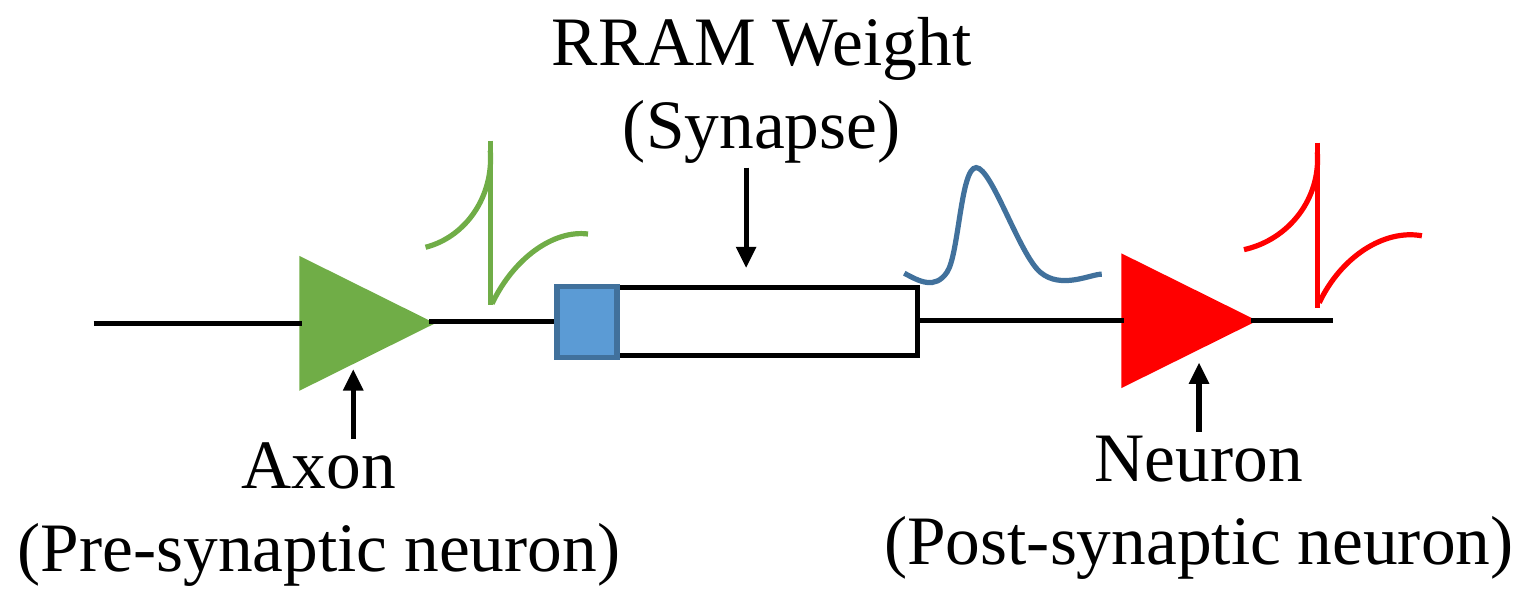}}
\caption{RRAM synapse between an axon and a neuron.}
\label{RRAM_synapse}
\end{figure}

RRAM is a two terminal non-volatile device with a conducting dielectric layer sandwiched between two electrodes as shown in fig. \ref{RRAM_synapse}. Electrically induced resistive switching effects shown in metal-insulator-metal systems are the basis for RRAM \cite{RRAM_nano}. Manipulation of oxygen vacancies in the conductance layer using positive and negative voltages helps in controlling current flow in RRAM. The state of RRAM reflects the current passed through it in the history, making it useful for modelling the synaptic weights of neurological synapses and implementing neural network architectures. In the case of spiking neural network (SNN), the tunable resistive state of RRAM synapses is analogous to the synaptic plasticity in brain. The electrical connection between a presynaptic neuron and a postsynaptic neuron (as shown in fig. \ref{RRAM_synapse}) changes, strengthening or weakening the synaptic impulses thus making it a case for brain-like pattern recognition.

\subsection{Crossbar array of RRAM Synapses}
\label{sec:1.3.2}

\begin{figure}[htbp]
\centerline{\includegraphics[scale=0.5]{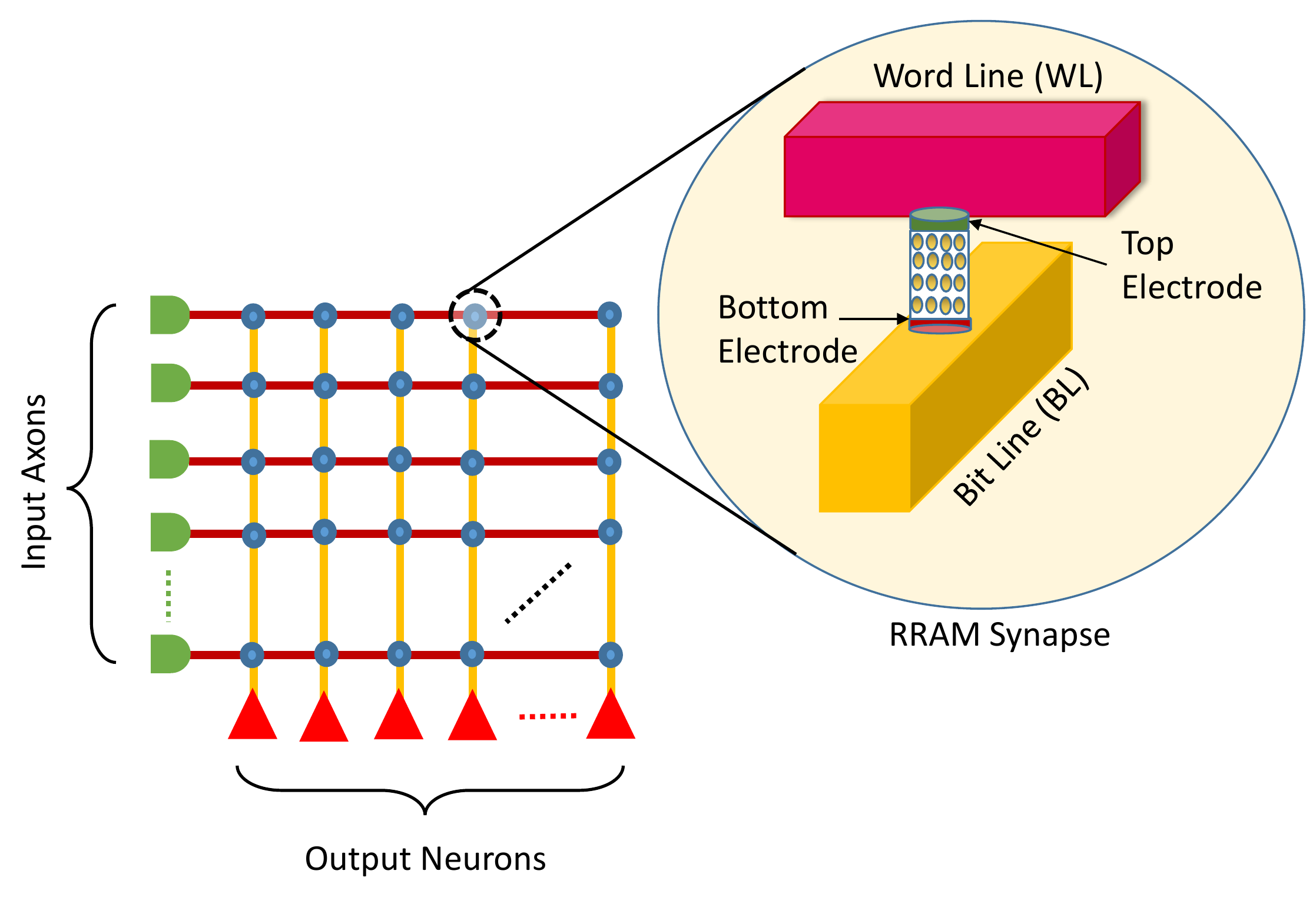}}
\caption{Crossbar array of synapses in a neuromorphic core.}
\label{Crossbar}
\end{figure} 
A crossbar architecture is shown in Fig. \ref{Crossbar}. Input axons are the input connections from the output neurons in the previous convolution layer which was mapped onto another neuromorphic core. Output neurons are spiking neurons. Spiking neurons recieve input current from many other spiking neurons (input axons as per the figure) and fire a spike when the integrated current input reaches the neuron threshold. These building blocks like axons, neurons and synapses together can perform mathematical operations. Matrix dot vector multiplcations can be performed efficiently with these crossbar structure \cite{MVM}. Each column in a crossbar produces the sum of product of input from axons and the weights stored at each RRAM synapses.

\subsection{Computation in a crossbar array of RRAM synapses}
\label{sec:1.3.3}

\begin{figure}[htbp]
\centerline{\includegraphics[scale=0.5]{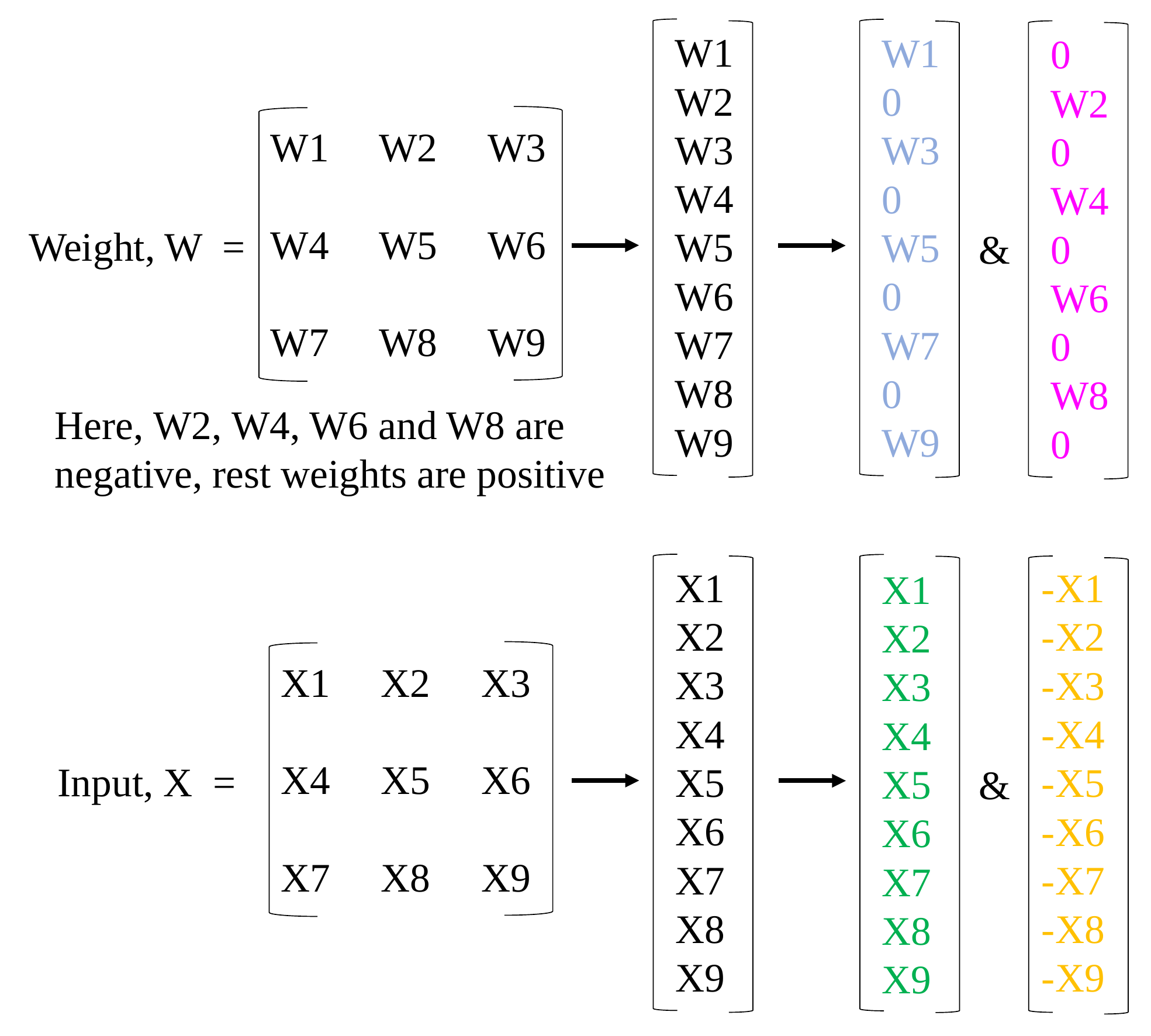}}
\caption{The weights and input activations used in the crossbar architecture of a neuromorphic core. Weights and inputs marked in different color is corresponding to its sign as marked with the similar color in fig. \ref{Crossbar}. }
\label{Crossbar_weights}
\end{figure}

\begin{figure}[htbp]
\centerline{\includegraphics[scale=0.6]{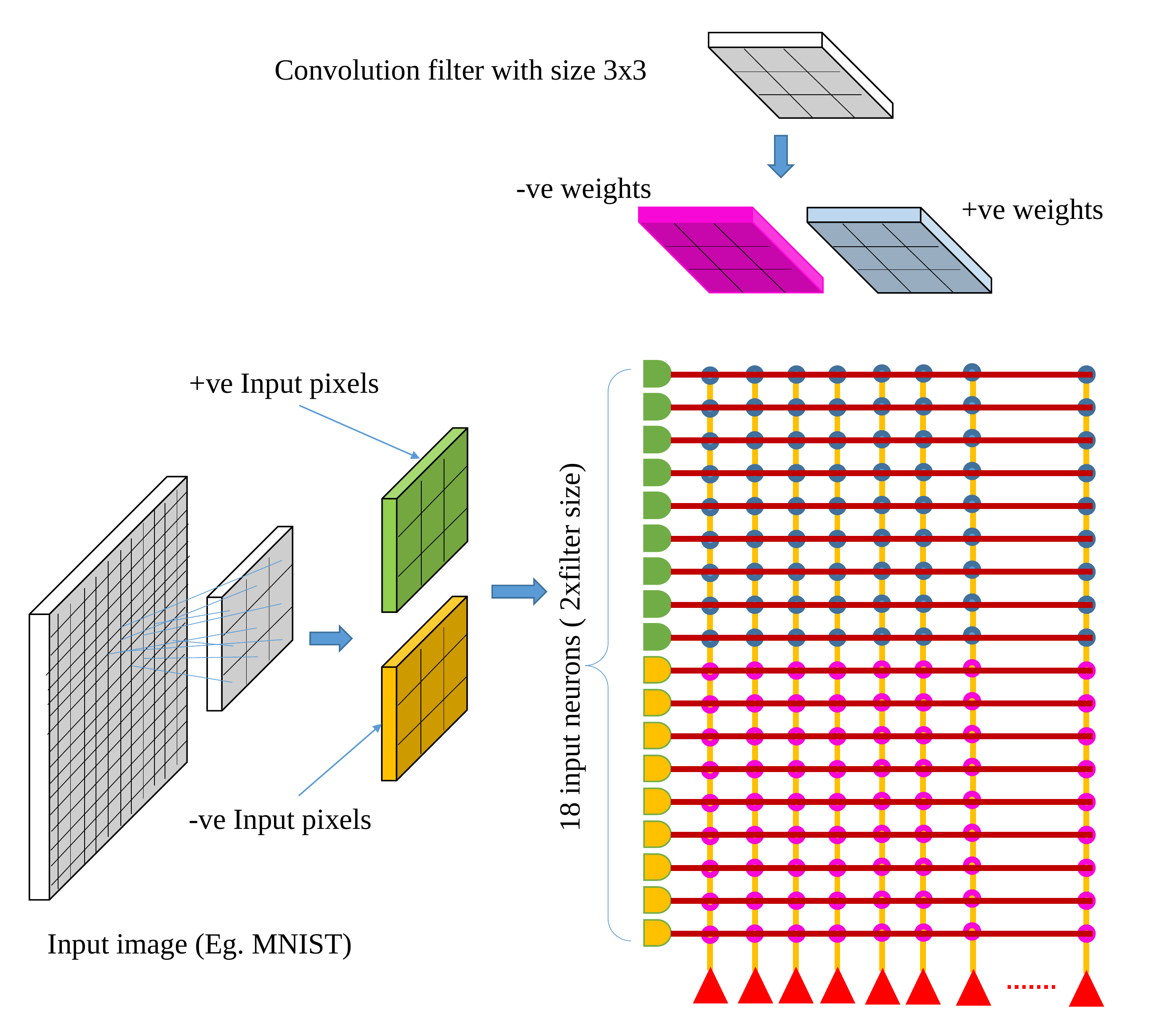}}
\caption{Illustration of computation in a crossbar array of synapses in a neuromorphic core.}
\label{Crossbar}
\end{figure}

The convolution operation in a convolutional neural network is implemented with the help of crossbar array of synapses in a neuromorphic core. Suppose a 3x3 convolution filter kernel after training a convolutional neural network is saved as weights, W as shown in fig. \ref{Crossbar_weights}. Out of these weights W2, W4, W6 and W8 are negative weights and rest of the weights are positive. Thus for the implementation of convolution operation this weight kernel W has to be separated into positive weights (marked in blue) and negative weights (marked in pink). Similarly for the input, X, it is divided into two matrices one is positive (marked in green) and another is negated input, -X (marked in orange). Once the weight matrix and input matrix is ready, the weight matrix can be written into the RRAM synapses -- positive weights occupy the top of the crossbar column while negative weights occupy the bottom part, while the inputs are also respectively fed into axons -- positive inputs are given to the top part of the axons and negative inputs to the bottom part of the axons. The convolution operation in a crossbar array between the inputs and the weights kernel is explicitly illustrated in fig. \ref{Crossbar} as mentioned in \cite{Alom_2016}. One of the disadvantage of such implementation is the utilization of double the amount of input axons (2 x filter size) needed as well as double the number of RRAM synapses. Half of the RRAM synapses has to be written with low conductance state. This work is also extended to make the architecture extremely parallel by stretching the separated weight matrices as in the toeplitz matrix \cite{Alom_2017}. But, the same disadvantages of poor utilization of axons and synapses as mentioned above will remain.    A slightly different approach of implementation is utilized in IBM's truenorth chip \cite{IBM_Esser}. They have only ternary weights (-1, 0, +1) and uses two crossbar synapses in a column as a single synapse to implement ternary weights. This will also end up using double the amount of physical synapses on neuromorphic chip compared to actual number of synapses in a weight kernel. Hence, truenorth also has a disadvantage of poor utilization of axons and synapses. Truenorth's actual physical core size, meaning number of axons X number of neurons, is 256 X 256, but literally their core size is only 128 X 256 to implement ternary weights.        

\subsection{Deep Neural Network (DNN) architecture}
\label{sec:1.3.4}

\begin{figure}[htbp]
\centerline{\includegraphics[scale=0.75]{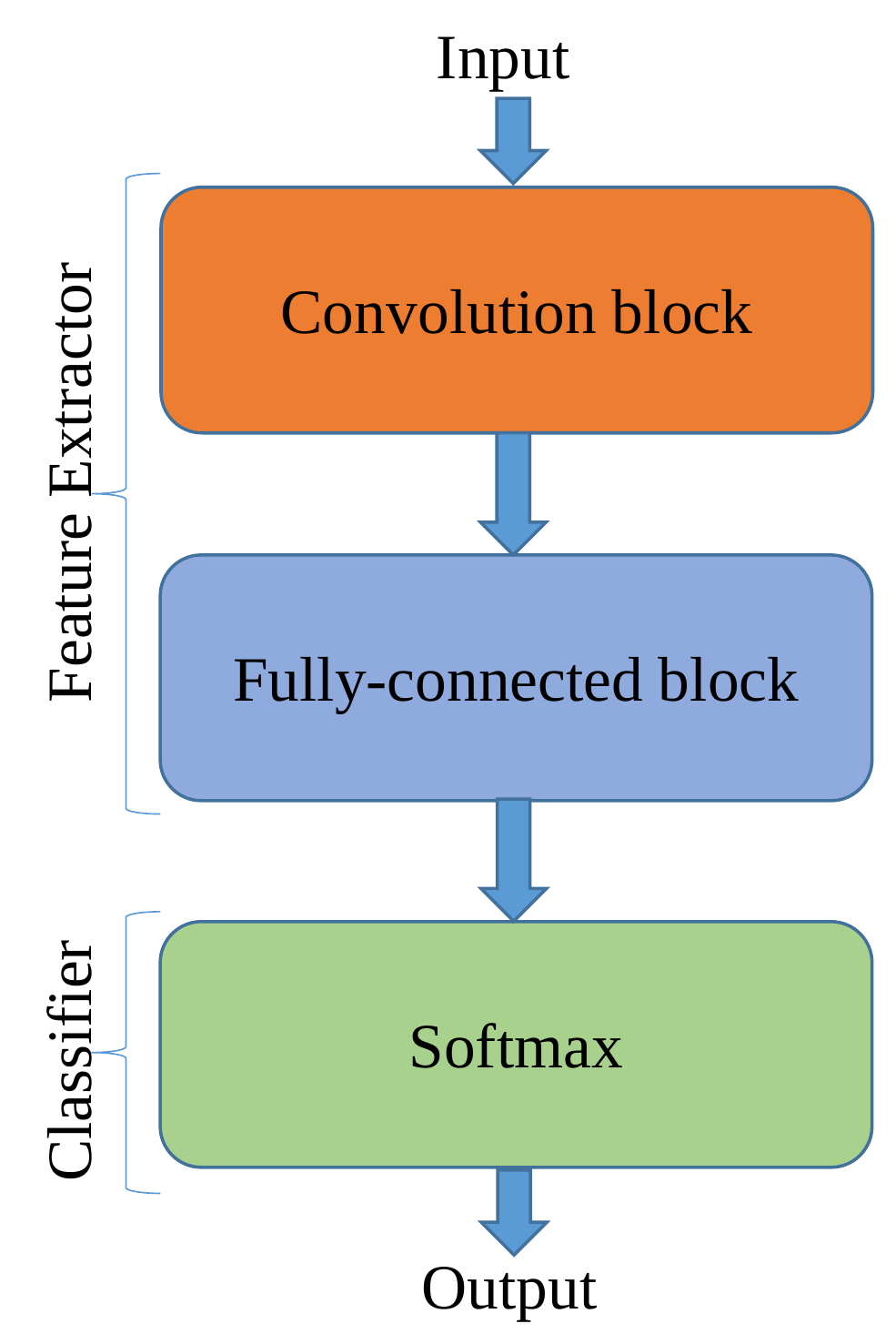}}
\caption{\textbf{Block diagram of a conventional deep neural network architecture (the convolutional neural network).} It comprises of mainly three blocks: the convolutional block, the fully connected block and the softmax layer.}
\label{fig:dnn_block}
\end{figure}

Deep learning has made much progress in recent years so much so that it has even outperformed humans in certain tasks, for instance, beating the current GO world champion \cite{Alphago}. DNN or deep CNN has achieved state-of-the-art accuracy in many image classification or patten recognition tasks such as handwritten digit recognition \cite{MNIST}, and several other datasets such as CIFAR \cite{CIFAR}, and ImageNet \cite{IMAGENET}. However, these networks typically need large amount of labeled training data; ImageNet has over 1 million labeled images for training.

A conventional CNN is shown in figure \ref{fig:dnn_block}. It comprises of mainly three blocks: the first block is made up of convolution layers, the second of fully connected layers and the third is the softmax layer. The convolution block contains convolution layers that perform the convolution operation on intermediate output activations. The convolution block also contains other layers that perform batch normalization or pooling. The fully connected block contains several layers of fully connected neural network. These two blocks are mainly for feature extraction. The final layer is a fully connected classifier which gives an output based on the softmax function. A typical learning algorithm used in a CNN is backpropagation of errors with stochastic gradient descent. The network parameters such as weights and biases are adjusted during training so as to predict the object label of an input image during testing.

\subsection{Spiking Neural Network (SNN) architecture}
\label{sec:1.3.5}

\begin{figure}[htbp]
\centerline{\includegraphics[scale=.6]{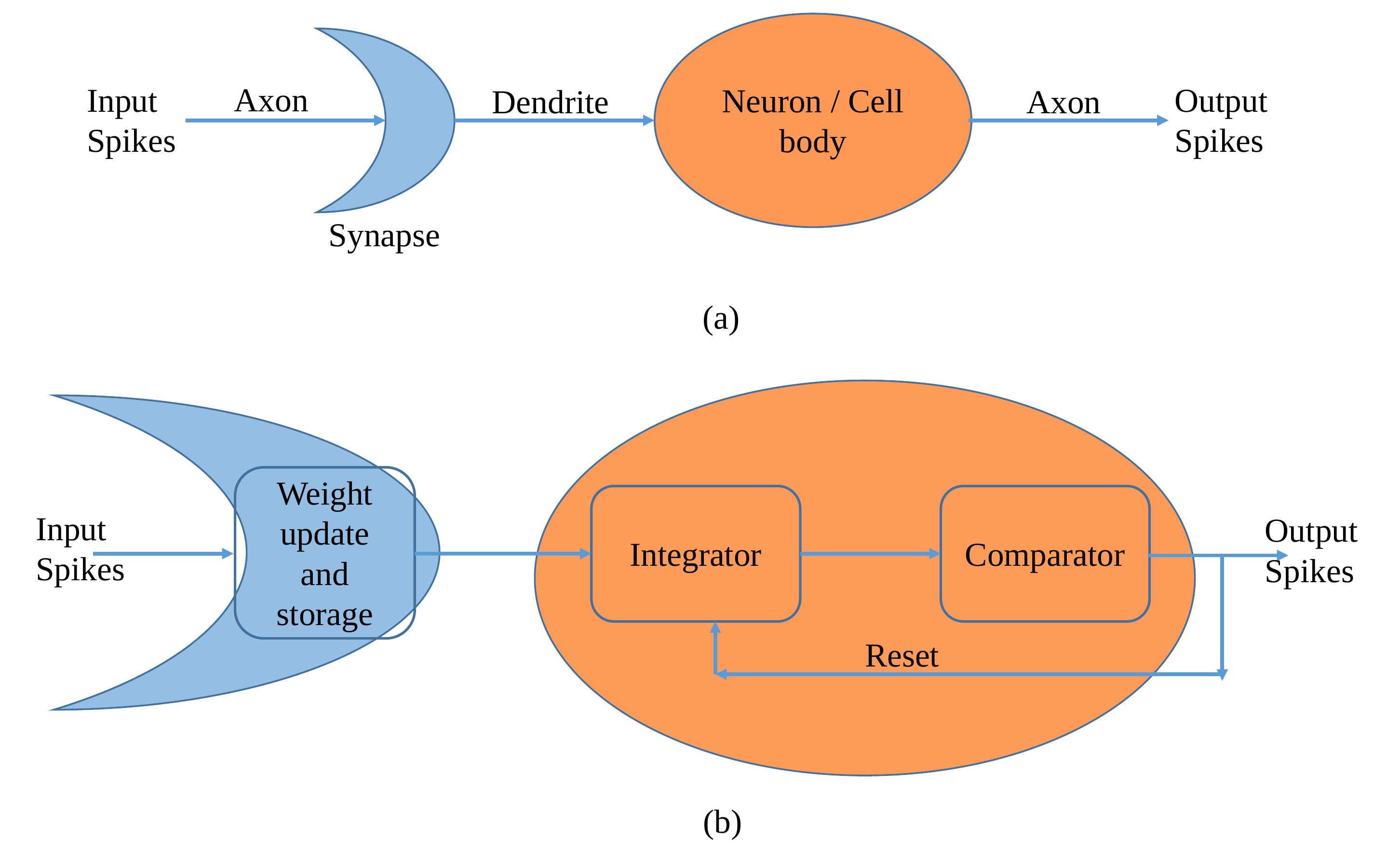}}
\caption{\textbf{Block diagram of a neuron in a SNN:} implemented in blocks as shown, namely, the synapse and the neuron.}
\label{fig:snn_block}
\end{figure}

Spiking neural network (SNN) is considered as the third generation of neural networks \cite{Maass}. SNN is inspired by biological neural networks while the DNN less so; hence the DNN is also commonly referred to as artificial neural networks (ANN). DNN does not have any biological roots apart from the hierarchical structure it possess \cite{Neocognitron}. SNN is event based: neural activations are communicated through spikes. Spiking neurons integrate incoming input spikes and emit a spike which is a threshold crossing event, as and when new information needs to be processed or communicated. These spikes are communicated through synapses which are associated with a weight quantity. 

A neuron in a SNN and its hardware implementation is shown in figure \ref{fig:snn_block}. The above figure \ref{fig:snn_block} (a) shows a single neuron (as part of a SNN) with its input and output mechanisms. The synapse is the connection between the axon of a pre-synaptic neuron and the dendrite of a post-synaptic neuron. A neuron integrates the incoming spikes received through its dendrites and may then emit a spike in the event of threshold crossing through its axon to its post-synaptic neurons. Figure \ref{fig:snn_block} (b) shows a block diagram representation of the biological model as in figure \ref{fig:snn_block} (a). The synapse is a storage element with input spikes and output current. Neuron computation is done using an integrator and a comparator. The integrator accumulates the input currents in terms of potential difference, which emulates the membrane potential in biological neurons. The comparator then checks if the membrane potential crosses the voltage threshold; a spike is emitted if crossed and the membrane potential is then reset to its baseline value.   

\subsection{Conversion of DNN to the spike-based domain: Spiking Deep Neural Network (SDNN)}
\label{sec:1.3.6}

In a conventional CPU or GPU, it requires more time and energy to run a SDNN, whereas the power consumption and computational latency in neuromorphic analog or digital dedicated hardwares \cite{Neurogrid, Spinnaker, IBM-truenorth} are orders of magnitude less. The substantial computational cost incurred during training and inference in a deep network for real world practical applications has created \cite{Peter_2015} :
\begin{itemize}
\item a need for specialized hardware acceleration.
\item a new computational paradigm.
\end{itemize}
One emerging approach is to convert the pre-trained DNN into SNN (while retaining its parameters) so that it can be mapped directly onto a neuromorphic hardware with little performance loss.

The spike based computation in the SNN consumes much less power compared to the high precision digital computation in the DNN. DNN has better classification accuracy compared to SNN. Hence, mapping a deep CNN to a SDNN potentially allows us to achieve better accuracy with high energy efficiency.
\begin{equation}
\begin{split}
& Spiking \; Neural \; Network => Low \; Energy \\
& Deep \; Neural \; Network => Better \; Accuracy \\
& SDNN => Better \; Accuracy \; Low \; Energy \; (BALE)\; Neural\; Network
\end{split}
\end{equation}
While it is difficult to achieve in a mapped SDNN the same level of accuracy as the DNN, research is ongoing to develop better mapping techniques.

\subsubsection{DNN to SNN conversion: SDNN background}
\label{sec:1.3.6.1}

DNN to SNN conversion techniques were developed in the ongoing research to map a trained neural network in conventional frame-based vision system representation to an event-based one \cite{Perez_2013}. Neurons in the frame-based CNN were converted to event-based neurons with leak, membrane potential reset and refractory periods.

One of the first research paper on CNN to SNN conversion is \cite{Cao_2015}. The conventional CNN is first converted into a tailored CNN which fulfils the requirements of the SNN. This tailored CNN is then trained. Finally, this tailored CNN is converted into a spiking CNN, while retaining the trained weights. The requirements imposed by the SNN on the tailored CNN are 
\begin{itemize}
\item using RELU \cite{Alex_2012} as activation functions,
\item removing biases from convolution and fully connected layers and
\item using spatial linear subsampling in place of maxpooling.
\end{itemize}

\cite{Peter_2015} extended the work of \cite{Cao_2015} by adding weight normalization techniques to improve the conversion accuracy. The approximation errors in SNNs due to either excessive or too little spikes are avoided by rescaling of weights. Model based and data based weight normalization techniques were proposed; data based normalization gives no loss in conversion accuracy for classification of MNIST dataset.

The integrate and fire (IF) neuron model was extensively used in SDNN until \cite{Huns_2015} demonstrated that a CNN can also be mapped onto a SDNN made up of leaky integrate and fire (LIF) neurons which are more biological plausible. This is achieved by using a modified LIF neuron known as the softened LIF neuron and by training the network with noise so as to improve network robustness against the variability inherent in spikes.

The hardware constrained neuromorphic algorithm is implemented in \cite{IBM_Esser} on the IBM Truenorth neuromorphic chip. The hardware constraints are namely, low precision weights and restricted connectivity among spiking neurons.

Adapting SNN is introduced in \cite{Bohte_2016}, which is based on adaptive spiking neurons. Asynchronous pulsed sigma-delta coding scheme is used by these spiking neurons to efficiently encode information in spike trains, while homeostatically optimizing the firing rate. This method uses an order of magnitude less spikes compared to other SDNN approaches; the RELU neurons in an ANN could be directly mapped to adaptive spiking neurons during conversion.

\subsubsection{General steps for conversion}
\label{sec:1.3.6.2}

The conversion of a pre-trained DNN to the event-based domain is for inference purposes. The principle of the conversion technique as mentioned in \cite{Cao_2015} is that the time averaged firing rate of  a spiking neuron must be correlated with the activation value of the corresponding neuron in the ANN. The generic steps involved for network conversion is as mentioned below:
\begin{itemize}
\item Choose a CNN to train.
\item Use ReLU for activation functions in the CNN.
\item Fix the bias to zero throughout training using stochastic gradient descent.
\item Save all the weights after training.
\item Replace neurons in the CNN with integrate and fire neurons without refractory period.
\item Map the saved weights to the SNN.
\item Convert the input image to poisson spike trains with firing rates proportional to each pixel intensity value.
\end{itemize}

\subsubsection{Factors affecting conversion accuracy}
\label{sec:1.3.6.3}

The issues affecting conversion accuracy as mentioned in \cite{Cao_2015} are: in the CNN the weights and biases can be negative. Since input integration is a weighted sum of inputs and the bias, the output can be negative. If the sigmoid function is used for activation it may also be negative. It is difficult to represent negative activations in the CNN on a SNN. It is also difficult to represent biases in the SNN. Two layer neural network is needed to implement spatial maxpooling in the SNN.

CNN to SNN mapping requires the input image to be converted to poisson spike trains with firing rates proportional to the pixel intensity value. As a result, the loss of accuracy during conversion can happen due to the factors \cite{Peter_2015}: Input spikes are not enough to result in threshold crossing, hence no output spike is emitted when activation values in the CNN are below threshold. If the spiking neuron receives too many input spikes in a single timestep or if some of its synaptic weights are higher than threshold, then the spiking neuron should emit more than one spike per timestep, which it cannot, and hence introducing error in the process. Due to the non-uniformity of the spike trains or the stochastic nature of the spiking input, a specific feature set could be over- or under- activated by incoming spikes.

An analysis of conversion and its theory is proposed in \cite{Bodo_2016}. One on one mapping of the spiking neuron and the activation function of the CNN, reveals that during threshold crossing, the membrane potential reached maybe of any value above threshold. This error would accumulate over time.

\subsubsection{Solution to the issues affecting conversion accuracy}
\label{sec:1.3.6.4}
The solution to the above-mentioned issues are the following (listed as above):
1. as mentioned in \cite{Cao_2015}, are to remove biases from convolution layers, use ReLU as activation function and use spatial linear subsampling instead of maxpooling.
2. as mentioned in \cite{Peter_2015} use weight normalization.
3. as mentioned in \cite{Bodo_2016} use reset by subtraction instead of reset to zero for spiking neurons. Instead of removing biases from convolutional layers, a constant input current can be applied to emulate the biases. Also apply normalization techniques.
4. as mentioned in \cite{Bohte_2016}, to reduce the variability of input spikes, the multi-bit values of the input maybe fed directly into the first hidden layer and spikes are then output henceforth.
5. as mentioned in \cite{Pool}, pooling layers can be avoided in deep neural network. Hence, even though there are techniques to convert pooling layers in SDNN, we can remove these layers from the DNN for simplicity sake.

\subsection{Spike based algorithms}
\label{sec:1.3.4}

In the past decade, spike timing dependent plasticity (STDP) has been a popular unsupervised learning method due to its biological plausibility \cite{Bi&Poo, Zhang_1998, Abbott_2000}. STDP mechanism depends on the timing difference between the pre-synaptic and post-synaptic spikes to adjust the synaptic weight. In the simple, doublet STDP \cite{Shubha, D-STDP_Roshan, D-STDP_Roshan_TNNLS}, when a post-synaptic spike happens after a pre-synaptic spike has arrived (pre-post event), then the weight of the synapse increases i.e. synaptic potentiation takes place; whereas, if a post-synaptic spike happens before a pre-synaptic spike (post-pre event), then the weight of the synapse decreases, i.e. depotentiation takes place. Similar to the doublet STDP, there is another variant of STDP called the triplet STDP \cite{Pfister_2006, Mostafa, T-STDP_Roshan, T-STDP_Roshan_ISCAS},  whereby, three spike events are considered (pre-post-pre, post-pre-post etc.). 

There are CMOS devices such as the floating gate MOSFET or nano-technology devices such as the memristors, Resistive Random Access Memories (RRAM), Phase Change Memories (PCM) and Spin-Transfer Torque Magnetic Random Access Memories (STT-MRAMs) used for the implementation of artificial synapses. One of the challenge is to integrate these nano-technology devices with CMOS. The characteristics of high synaptic density on neuromorphic hardware has to be compromised. Understanding the device physics becomes the key for the implementation of artificial synapses, especially while using any of the technologies such as floating gate MOSFET, memristors or the more recent spin devices to implement plasticity rules such as STDP.


\subsection{Conclusion}
\label{sec:1.4}



For future work, it maybe worth investigating conversion of DNN using different encoding schemes such as temporal coding or latency coding instead of just rate coding. This would reduce the number of spikes required to represent an input and result in more efficient computing. In the long run however, hardware compatible SNN algorithms should be developed that enable on-chip learning and inference for various applications. This will eliminate the need for conversion of DNN to SNN; the challenge would be how one may improve the accuracy of such SNN algorithms. 

Here, we have given an overview of the current state-of-art neuromorphic algorithms on RRAM based neuromorphic devices. While research is on-going to develop SNN for on-chip learning, the current reliable approach for a real-world application is to do inferencing on-chip based on a converted DNN that is pre-trained off-chip. During the mapping of a DNN to neuromorphic hardware, hardware constraints such as number of neurons and synapses, core size, fan in-fan out degrees, routing, spike traffic congestion etc. have to be taken into consideration. Should any of the above constraints not be met, the DNN architecture will have to be modified accordingly, so as to fit into a specific neuromorphic hardware. Given its small form factor and energy efficiency, neuromorphic hardware is well suited for edge computing applications in the fields of robotics, surveillance, unmanned aerial vehicles etc.
 
\section*{Funding}
This research is supported by Programmatic grant no. A1687b0033 from the Singapore governments Research, Innovation and Enterprise 2020 plan (Advanced Manufacturing and Engineering domain).


%
%
%

\end{document}